\documentclass[runningheads]{llncs}
\usepackage{graphicx}
\usepackage{amsmath}
\usepackage{amssymb}
\usepackage[english]{babel}
\usepackage{proof}
\usepackage{xspace}

\usepackage{ebproof}	%prooftrees
\usepackage{graphicx}   %scalebox

\usepackage{cancel}
\usepackage{xcolor}

\usepackage{wasysym} %emojis
 
\usepackage{tikz}
\usetikzlibrary{arrows, positioning, calc}
\usetikzlibrary{automata,positioning}

\usepackage{multicol}
\usepackage{longtable}

\usepackage[toc,page]{appendix}

\usepackage{hyperref}

\usepackage{comment}

\newcommand{\M}{\mathcal{M}}
\newcommand{\W}{\mathsf{W}}
\newcommand{\R}{\mathsf{R}}
\newcommand{\V}{\mathsf{V}}
\newcommand{\g}{\mathsf{g}}
\newcommand{\w}{\mathsf{w}}
\renewcommand{\v}{\mathsf{v}}

\newcommand{\I}{\emph{I}\xspace}
\newcommand{\You}{\emph{You}\xspace}
\newcommand{\My}{\emph{My}\xspace}

\newcommand{\Me}{\emph{Me}\xspace}
\newcommand{\Your}{\emph{Your}\xspace}

\tikzset{
modal/.style={>=stealth’,shorten >=1pt,shorten <=1pt,auto,node distance=1.5cm,
semithick},
world/.style={circle,draw,minimum size=0.5cm,fill=gray!15},
point/.style={circle,draw,inner sep=0.5mm,fill=black},
reflexive above/.style={->,loop,looseness=7,in=120,out=60},
reflexive below/.style={->,loop,looseness=7,in=240,out=300},
reflexive left/.style={->,loop,looseness=7,in=150,out=210},
reflexive right/.style={->,loop,looseness=7,in=30,out=330}
}

\usepackage{grid-system}

\allowdisplaybreaks

\begin{document}
\title {\texorpdfstring{Games for Hybrid Logic \thanks{Research supported by FWF projects P 32684 and W 1255.}\\ {\large From Semantic Games to Analytic Calculi}}{}}

\titlerunning{Games for Hybrid Logic}

\author{\texorpdfstring{Robert Freiman\inst{1}\orcidID{0000-0001-8251-4272}}{}}

\authorrunning{R. Freiman}

\institute{\texorpdfstring{Institute of Logic and Computation, TU Wien, Vienna, Austria \\ \email{robert@logic.at}}{}}

\maketitle

\begin{abstract}
Game semantics and winning strategies offer a potential  conceptual bridge between semantics and proof systems of logics. We illustrate this link for hybrid logic -- an extension of modal logic that allows for explicit reference to worlds within the language. The main result is that the systematic search of winning strategies over all models can be finitized and thus reformulated as a proof system.

\keywords{\texorpdfstring{Hybrid logic \and games \and proof systems.}{Hybrid logic, games, proof}}
\end{abstract}

\section{Introduction}
Games have a long and varied tradition in logic (see, for example \cite{sep-logic-games,vbenthbook,vbook}). Building on the concepts of rational behavior and strategic thinking, they offer a fruitful natural approach to logic, complementing the common paradigm of model-theoretic semantics and proof systems. Game semantics goes back to Jaakko Hintikka \cite{Hintikka1973-HINLLA}, who designed a game for two players, usually called \Me (or \I) and \You, seeking to establish the truth of a formula $\phi$ of first-order logic in a model $\mathcal{M}$. The game proceeds by rules for step-wise reducing $\phi$ to an atomic formula. The winning condition depends on the truth of this atomic formula in $\mathcal{M}$. It turns out that \I have a winning strategy for this game if and only if $\phi$ is true in $\mathcal{M}$. A natural question is, whether there is an algorithm for searching for winning strategies for the game of $\phi$ over \emph{all} models, which could thus establish (or refute) the validity $\phi$. One such approach are \emph{disjunctive winning strategies} allowing the players to keep track and -- if necessary -- revise their choices, depending on the truth values of the atomic sentences in the current model. This technique has been first demonstrated in \cite{FMGiles} for Giles' game for Łukasiewicz logic.

In this paper, we apply this method to the case of hybrid logic. Hybrid logic is an extension of the possible-world semantics of ``orthodox" modal logic allowing one to explicitly refer to worlds within the object language, using nominals, while keeping many attractive features of modal logic intact. Obviously, this increased expressivity  allows us to get a grip on many frame properties that are provably not expressible in orthodox modal logic  \cite{blackburn2002modal}. Apart from this, using nominals can be an advantage for modelling in temporal logic \cite{hybltemp} or in making the link between modal logic and description logics more explicit \cite{hybconclang}. The choice of hybrid logic over orthodox modal logic for the development of a game-theoretic approach is a natural one: Explicit reference of particular worlds within the language provides conceptual clarity for the lifting of the semantic game to disjunctive strategies. 

The coherence of hybrid logic with games has been demonstrated by Patrick Blackburn, who designed a Lorenzen-style \cite{lorenzen} dialogue game for hybrid logic \cite{blackburndialog}. Sara Negri presented the labeled proof-system $\mathbf{G3K}$ that resembles the semantics of modal logic \cite{NegriKripke}. Our approach brings together the best of two worlds: it supplements the clear semantic motivation of $\mathbf{G3K}$ with an accessible game-theoretic viewpoint of Blackburn's dialogue game with a direct semantic link. Similar to $\mathbf{G3K}$, a failed search for a disjunctive winning strategy directly gives rise to a countermodel. 

The main result of this paper addresses the following difficulty: generally, while searching for a disjunctive winning strategy, one has to keep track of all infinitely many possible models and (in the case of modal operators) infinitely many possible choices. The perhaps surprising and highly non-trivial insight is, that this search can be finitized in a way that is coherent with respect to the semantics-provability-bridge mentioned above. This is achieved by a conceptual reduction of choices of the opponent to an \emph{optimal choice}. Thus, the search for a disjunctive winning strategy can itself be formulated as a proof-system.

This paper is structured as follows: Section~\ref{SectHybrlog} is a recap of hybrid logic. In Section~\ref{game} we present the semantic game over a model and formalize strategies. The disjunctive game is introduced in Section~\ref{disjgame}, which also contains the main result.  Finally, its finitized formulation as an analytic calculus is presented in Section~\ref{proofsys}.

\section{Preliminaries}\label{SectHybrlog}
The language of hybrid modal logic is as follows: We start from two disjoint, countably infinite sets $N$ (set of nominals) and $P$ (set of propositional variables). Nominals are usually called ``$i$, $j$, $k$, \ldots'' propositional variables are called ``$p$, $q$, \ldots " Formulas $\phi$ are built according to the following grammar:

\[ \phi ::=
p\mid 
i\mid
R(i,j)\mid
\phi \wedge \phi\mid
\phi \vee \phi\mid
\phi \rightarrow \phi\mid
\neg \phi
\mid @_i \phi
\mid \Box \phi
\mid\Diamond \phi\]
Formulas of the form $p$, $i$ and $R(i,j)$ are called \emph{elementary}. Intuitively, we can think of the nominal $i$ as being the name of a particular world in a model. Hence, $i$ is true in exactly one world. The formula $@_i \phi$ stands for the fact that $\phi$ is true in the world with name $i$. The relational claim $R(i,j)$ says that the world with name $j$ is accessible from the world with name $i$. It is usually defined as $@_i \Diamond j$. However, we leave it as an elementary formula in order to prevent circular definitions in the description of the game rules in the following sections.

We now define the semantics formally: A model $\mathcal{M}$ for hybrid modal logic is a tuple $(\W,\R,\V,\g)$, where

\begin{enumerate}
\item $\W$ is a non-empty set. Its elements are called \emph{worlds}.
\item $\R \subseteq \W\times \W$ is called \emph{accessibility relation}. As usual, we write $\w\R\v$ instead of $(\w,\v)\in \R$. The set of \emph{accessible worlds from} $\w$ are  $\w\R := \{\v\in \W: \w\R\v\}$.
\item $\V: P \rightarrow \mathcal{P}(\W)$\footnote{$\mathcal{P}(\W)$ denotes the power set of $\W$.} is called \emph{valuation function}.
\item $\g: N\rightarrow \W$ is called \emph{assignment}. If $\g(i)=\w$, we say that $i$ is a \emph{name} of $\w$, or simply that $\w$ has a name.
\end{enumerate}

Truth of formulas in the world $\w$ of $\W$ is defined recursively:
\begin{align*}
\M, \w & \models p \text{, iff } \w\in \V(p),\\
\M, \w & \models i \text{, iff } \g(i) =\w,\\
\M, \w & \models R(i,j) \text{, iff } \g(i)\R\g(j),\\
\M,\w & \models \phi \wedge \psi \text{, iff } \M,\w\models \phi \text{ and } \M,\w \models \psi,\\
\M,\w & \models \phi \vee \psi \text{, iff } \M,\w\models \phi \text{ or } \M,\w \models \psi,\\
\M,\w & \models \phi \rightarrow \psi \text{, iff } \M,\w\not \models \phi \text{ or\footnotemark } \M,\w \models \psi,\\
\M,\w & \models \neg\phi \text{, iff } \M,\w \not \models \phi,\\
\M,\w & \models @_i\phi \text{, iff } \M,\g(i) \models \phi,\\
\M,\w & \models \Box \phi \text{, iff for all\footnotemark } \v\in \W, (\w,\v) \notin \R \text{ or } \M,\v  \models \phi,\\
\M,\w & \models \Diamond \phi \text{, iff for some } \v\in \W, \w\R\v \text{ and } \M,\v  \models \phi.\\
\end{align*}

\addtocounter{footnote}{-1}
\footnotetext{This is equivalent to the usual ``if $\M, \w \models \phi$, then $\M,\w \models \psi$".}
\stepcounter{footnote}
\footnotetext{This is equivalent to the usual ``For all $\v \in \W$, if $\w\R\v$, then $\M,\v \models \phi$". Our formulation is more easily representable as a game rule.}
\vspace{-5mm}
We say that a formula $\phi$ is true in the model $\M$ and write $\M \models \phi$ iff for every world $\w$, $\M, \w \models \phi$. For a class $\mathfrak{M}$ of models, we write $\mathfrak{M}\models \phi$  and say that $\phi$ is \emph{valid over} $\mathfrak{M}$ iff for all $\M \in \mathfrak{M}$, $\M \models \phi$. We say that $\phi$ is \emph{valid} (we write $\models \phi$) iff $\phi$  is valid over the class of all models.

The most important class for what follows is the class of named models $\mathfrak{N}$. A model is called \emph{named} iff the assignment is surjective, i.e. every world has a name. The accessibility relation in such named models is completely determined by the truth value of the relational formulas. For these models, we can thus state the following semantic facts about the modal operators without explicitly referring to the semantics of the accessibility relation. Let $\w = \g(i)$, then
\begin{align*}
\M,\w & \models \Box \phi \text{, iff for all } j\in N, \M,\g(j) \models \neg R(i,j) \vee \phi,\\
\M,\w & \models \Diamond \phi \text{, iff for some } j\in N, \M,\g(j) \models R(i,j) \wedge \phi.\\
\M & \models \phi \text{, iff for all } i\in N, \M,\g(i) \models \phi.
\end{align*}
Named models are important, since questions of validity can be reduced to questions of validity over the class of named models:
\begin{lemma}[Theorem 7.29 in \cite{blackburn2002modal}] \label{namedlemma}
$\models \phi \iff \mathfrak{N} \models \phi$.
\end{lemma}

\begin{comment}
\begin{example}\label{namedexample}
The single pure forumla $R(i,j)\wedge R(j,k) \rightarrow R(i,k)$  defines the class of transitive frames. Hence, we can reduce the question of validity over transitive frames to questions of entailment of the set $\{R(i,j)\wedge R(j,k)\rightarrow R(i,k)\mid i,j,k\in N\}$ for named models. Note, that writing the $@$-operator in front of the formulas is unnecessary in this case, since truh of $R$-formulas does not depend on the particular world it is evaluated at.
\end{example}
\end{comment}

\section{A Game for Truth}
\label{game}
The \emph{semantic game} is played over a model $\mathcal{M}=(\W,\R,\V,\g)$ by two players, \Me and \You, who argue about the truth of a formula $\phi$ at a world $\w$. At each stage of the game, one player acts in the role of a proponent, while the other one acts as opponent of the claim that a formula $\phi$ is true at the world $\w$. We represent the situation where \I am the proponent (and \You are the opponent) by the \emph{game state} $\mathbf{P}, \w:\phi$, and the situation where \I am the opponent (and \You are the proponent) by $\mathbf{O}, \w:\phi$. We  add another kind of game state of the form $\mathbf{P}:\phi$ or $\mathbf{O}:\phi$ representing the claim that $\phi$ is true in the whole model.\\
In order to completely lay down the rules of the game, we must define its \emph{game tree}, i.e. a tree whose nodes are game states and are labeled either ``I" or ``Y". Each run of the game is identified with a path through that game tree:  in nodes labelled ``I" that have children, it is \My choice to decide which of the children the game continues at. It is \Your choice in nodes labelled ``Y". Each run of a game ends in a leaf state: if it is labelled ``I", \I win and \You lose, if it is labelled ``Y", \I lose and \You win. We denote the tree rooted in the game state $g$ by $\mathbf{G}(\M,g)$\footnote{We often write $\mathbf{G}(g)$, if $\M$ is clear from context}.
Below, we give a recursive definition of $\mathbf{G}(g)$ by specifying its immediate subtrees. We give the rules only for the case where \I am in the role of proponent. The variants of the rules where \I am in the role of the opponent is obtained by switching ``I'' and ``Y'' and ``$\mathbf{P}$" and ``$\mathbf{O}$" in the respective rule\footnote{For example, for $(R_\rightarrow)$, if $g=\mathbf{O},\w:\psi_1 \rightarrow \psi_2$, then $g$ is labelled ``Y" and its immediate subtrees are $\mathbf{G}(\mathbf{P}, \w: \psi_1)$ and $\mathbf{G}(\mathbf{O}, \w: \psi_2)$.}. Note that the rules closely parallel the recursive definition of truth of the previous section:

\begin{description}
\item[$(R_\vee)$] If $g = \mathbf{P}, \w: \psi_1\vee \psi_2$, then $g$ is labelled ``I" and its immediate subtrees are $\mathbf{G}(\mathbf{P}, \w: \psi_1)$ and $\mathbf{G}(\mathbf{P}, \w: \psi_2)$.

\item[$(R_\wedge)$] If $g = \mathbf{P}, \w: \psi_1\wedge \psi_2$, then $g$ is labelled ``Y" and its immediate subtrees are $\mathbf{G}(\mathbf{P}, \w: \psi_1)$ and $\mathbf{G}(\mathbf{P}, \w: \psi_2)$. 

\item[$(R_\rightarrow)$] If $g = \mathbf{P}, \w: \psi_1\rightarrow \psi_2$, then $g$ is labelled ``I" and its immediate subtrees are $\mathbf{G}(\mathbf{O}, \w: \psi_1)$ and $\mathbf{G}(\mathbf{P}, \w: \psi_2)$.

\item[$(R_\neg)$] If $g = \mathbf{P}, \w: \neg \psi$, then $g$ is labelled ``I" and its immediate subtree is $\mathbf{G}(\mathbf{O},\w:\psi)$.

\item[$(R_@)$] If $g = \mathbf{P}, \w: @_i \psi$, then $g$ is labelled ``I" and its immediate subtree is $\mathbf{G}(\mathbf{P},\g(i):\psi)$.

\item[$(R_\Box)$] If $g = \mathbf{P}, \w: \Box \psi$ and $\w\R \ne \emptyset$, then $g$ is labelled ``Y" and its immediate subtrees are $\mathbf{G}(\mathbf{P},\v:\psi)$, where $\v$ ranges over $\w\R$. If $\w\R = \emptyset$, then $g$ is a leaf labelled ``I".

\item[$(R_\Diamond)$] If $g = \mathbf{P}, \w: \Diamond \psi$ and $\w\R \ne \emptyset$, then $g$ is labelled ``I" and its immediate subtrees are $\mathbf{G}(\mathbf{P},\v:\psi)$, where $\v$ ranges over $\w\R$. If $\w\R = \emptyset$, then $g$ is a leaf labelled ``Y".

\item[$(R_p)$] If $g = \mathbf{P}, \w: p$, then $g$ is a leaf. It is labelled ``I" iff $\w \in \V(p)$.

\item[$(R_i)$] If $g = \mathbf{P}, \w: i$,  then $g$ is a leaf. It is labelled ``I" iff $\g(i)= \w$.

\item[$(R_{R})$] If $g = \mathbf{P}, \w: R(i,j)$, then $g$ is a leaf. It is labelled ``I" iff $\g(i)\R\g(j)$.

\item[$(R_U)$] If $g = \mathbf{P} : \psi$, then $g$ is labelled ``Y" and its immediate subtrees are $\mathbf{G}(\mathbf{P}, \w: \psi)$, where $\w$ ranges over $\W$.
\end{description}

Since the degree of the involved formula strictly decreases with every child, game trees are always of finite height. This means that every run of the game lasts only finitely many rounds.

\begin{example}\label{simpleExample}
Consider the following model $\M$ with worlds $\w_1, ..., \w_5$ and the accessibility relation represented by arrows. We write $p$ inside the circle representing a world $\w$ iff $\w \in \V(p)$.

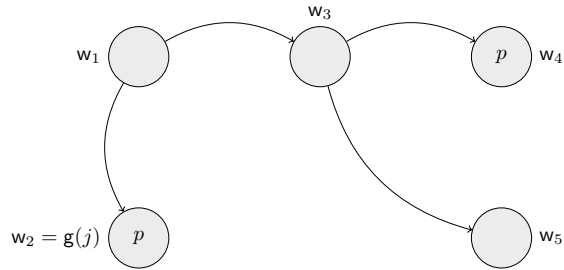
\begin{center}
\scalebox{0.8}{
\begin{tikzpicture}[node distance=2cm,world/.append style={minimum
size=1cm}]
\node[world] (w1) [label=left: $\w_1$] {};
\node[world] (w2) [label=left: {\(\w_2 =\g(j)\)},below = of w1] {$p$};
\node[world] (w3) [label=above: $\w_3$, right =of w1] {};
\node[world] (w4) [label=right:$\w_4$, right=of w3] {$p$};
\node[world] (w5) [label=right: $\w_5$,below=of w4] {};
\path[->] (w1) edge[bend right] (w2);
\path[->] (w1) edge[bend left] (w3);
\path[->] (w3) edge[bend left] (w4);
\path[->] (w3) edge[bend right] (w5);
\end{tikzpicture}
}
\end{center}

Let us consider the following run of the game $\mathbf{G}(\M,\mathbf{P}, \w_1:\Box (j \vee \neg \Box p))$. First, \You must choose a neighboring world. Since \You know that \I can defend $j$ at $\w_2$, let us say that \You choose $\w_3$ and \I must then defend $j\vee \neg \Box p$ at $\w_3$. Clearly, \I will choose the second disjunct. According to the rule of negation, now a role switch occurs: \I am now the Opponent and \You the Proponent of $\Box p$ at $\w_3$. Hence, \I must choose a neighboring world and \You must defend $p$ there. As \My choice is between the $p$-world $\w_4$ and the non-$p$-world $\w_5$, \I will choose $\w_5$ and win the game. We can represent this run of the game as the following path:

\begin{center}
{\begin{tikzpicture}[level distance =1cm]
\tikzstyle{level 1}=[sibling distance=80mm]
\tikzstyle{level 2}=[sibling distance=40mm]
 		 \node {\(\left[\mathbf{P}, \w_1: \Box (j \vee \neg \Box p)\right]^Y\)}
 			child {node {\(\left[\mathbf{P}, \w_3: (j \vee \neg \Box p)\right]^I\)}
            child {node {\(\left[\mathbf{P}, \w_3: \neg \Box p\right]^I\)}
            child {node {\(\left[\mathbf{O}, \w_3: \Box p\right]^Y\)}
            child {node {\(\left[\mathbf{O}, \w_5: p\right]^I\)}
               }}}};
		\end{tikzpicture}}
\end{center}

The whole game tree represents all possible choices of the two players at any point:

\begin{center}
{\begin{tikzpicture}[level distance = 1cm]
\tikzstyle{level 1}=[sibling distance=60mm]
\tikzstyle{level 2}=[sibling distance=30mm]
\tikzstyle{level 2}=[sibling distance=20mm]
 		 \node {\(\left[\mathbf{P}, \w_1: \Box (j \vee \neg \Box p)\right]^Y\)}
    	child {node {\(\left[\mathbf{P}, \w_2: (j \vee \neg \Box p)\right]^I\)}
        child {node {\(\left[{\mathbf{P}, \w_2: j }\right]^I\)}
        edge from parent node[left]{}}
        child {node {\(\left[\mathbf{P}, \w_2: \neg\Box p\right]^I\)}
         child {node {\(\left[\mathbf{O}, \w_2: \Box p\right]^Y\)}}
         edge from parent node[right]{}}
         edge from parent node[left]{}}
        child {node {\(\left[\mathbf{P}, \w_3: (j \vee \neg \Box p)\right]^I\)}
        child {node {\(\left[\mathbf{P}, \w_3: j\right]^Y\)}
        edge from parent node[left]{}}
      	child {node {\(\left[\mathbf{P}, \w_3: \neg \Box p\right]^I\)}
        child {node {\(\left[\mathbf{O}, \w_3: \Box p\right]^I\)}
        child {node {\(\left[\mathbf{O}, \w_4:  p\right]^Y\)}
        edge from parent node[left]{}}
        child {node {\(\left[{\mathbf{O}, \w_5:  p}\right]^I\)}
        edge from parent node[right]{}}
        edge from parent node[right]{}}
        edge from parent node[right]{}}
        edge from parent node[right]{}};
		\end{tikzpicture}}
\end{center}

\I cannot win in the state $\mathbf{O}, \w_2: \Box p$, because there are no neighbors of $\w_2$.  Although not all possible runs of a game end in a winning state for \Me, \I can make choices that will guarantee that the game will end in \My victory. Generally, if \I can always enforce a given game to end in a winning state for \Me, we say that \I have a \emph{winning strategy} for that game. We will make this notion more formal in the following subsection.
\end{example}

\subsection*{Winning Strategies}
To precisely describe the scenario where \I can enforce that the game ends in a winning state for \Me, we need the notion of a winning strategy:

\begin{definition}
A \emph{strategy}\footnote{This deviates from the standard, more general game-theoretic definition of a winning strategy. It is sufficient for our purposes.} for \Me for the game $\mathbf{G}(\M, g)$ is a subtree obtained by removing from the game tree all but one children from every non-leaf node labelled ``I". A strategy for \Me is \emph{winning} if all leaf states are labelled ``I". (Winning) strategies for \You are defined symmetrically.
\end{definition}

\begin{example}
Continuing the game from Example~\ref{simpleExample}, we can now make precise our observation that \I can make choices such that the game will always end in winning states for \Me. A winning strategy for \Me for $\mathbf{G}(\mathbf{P}, \w_1:\Box(j\vee \neg \Box p))$ can be found in Figure~\ref{figstrategy}.

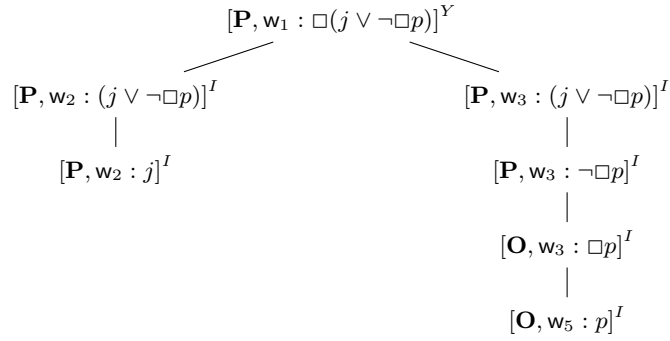
\begin{figure}
\begin{center}
{\begin{tikzpicture}[level distance = 1cm]
\tikzstyle{level 1}=[sibling distance=60mm]
\tikzstyle{level 2}=[sibling distance=40mm]
 		 \node {\(\left[\mathbf{P}, \w_1: \Box (j \vee \neg \Box p)\right]^Y\)}
    	child {node {\(\left[\mathbf{P}, \w_2: (j \vee \neg \Box p)\right]^I\)}
        child {node {\(\left[\mathbf{P}, \w_2: j\right]^I\)}}}
        child {node {\(\left[\mathbf{P}, \w_3: (j \vee \neg \Box p)\right]^I\)}
        child {node {\(\left[\mathbf{P}, \w_3: \neg \Box p\right]^I\)}
        child {node {\(\left[\mathbf{O}, \w_3:  \Box p\right]^I\)}
        child {node {\(\left[\mathbf{O}, \w_5:  p\right]^I\)}}}}};
		\end{tikzpicture}}
\end{center}
\caption{\label{figstrategy}A winning strategy for $\mathbf{G}(\mathbf{P}, \w_1:\Box(j\vee \neg \Box p))$}
\end{figure}

\end{example}

\begin{theorem}\label{winningstrategy}
Let $\mathcal{M}$ be a model and $\w$ a world.\footnote{Proofs can be found in the appendix.}
\begin{enumerate}
\item \I have a winning strategy for $\mathbf{G}(\M, \mathbf{P}, \w: \phi)$ iff $\mathcal{M},\w \models \phi$. 
\item \I have a winning strategy for $\mathbf{G}(\M, \mathbf{O}, \w: \phi)$ iff $\M,\w\not \models \phi$.
\item \I have a winning strategy for $\mathbf{G}(\M, \mathbf{P}: \phi)$ iff $\mathcal{M} \models \phi$.
\end{enumerate}
\end{theorem}

To prove the theorem we use the following basic lemma showing that having a winning strategy for $\mathbf{G}(\M, \mathbf{P}, \w: \phi)$ and $\mathbf{G}(\M, \mathbf{O}, \w: \phi)$ is exclusive. 

\begin{lemma}\label{winstratexclusivity}
\I have a winning strategy for $\mathbf{G}(\M,\mathbf{P}, \w:\phi)$ iff \I do not have a winning strategy for $\mathbf{G}(\M,\mathbf{O}, \w:\phi)$.
\end{lemma}

\begin{proof}
By induction on $\phi$.
\end{proof}

\begin{proof}[of Theorem~\ref{winningstrategy}]
2 follows from 1 in the obvious way. By Lemma~\ref{winstratexclusivity}, it suffices to show only 1. We show both directions simultaneously by induction on $\phi$. If $\phi$ is atomic, everything follows from the definition of a winning state. Let us show some of the inductive steps: if $\phi$ is of the form $\psi \wedge \chi$, then \I have a winning strategy  for $\mathbf{G}(\M,\mathbf{P},\w:\psi \wedge \chi)$ iff \I have winning strategies for both $\mathbf{G}(\M,\mathbf{P},\w:\psi)$ and $\mathbf{G}(\M,\mathbf{P},\w:\chi)$. By induction hypothesis, this is the case iff $\M, \w \models \psi$ and $\M,\w \models \chi$, which in turn is equivalent to $\M,\w\models \psi \wedge \chi$.\\
As for modal operators, \I have a winning strategy for $\mathbf{G}(\M,\mathbf{P},\w:\Box\psi)$ iff for all successors $\v$ of $\w$, \I have a winning strategy for  $\mathbf{G}(\M,\mathbf{P},\v:\psi)$. By induction hypothesis, this is the case iff $\M, \v \models \psi$. Since $\v$ is an arbitrary successor of $\w$, this is equivalent to $\M, \w \models \Box \psi$. Note that this also covers the trivial case, where there are no successors of $\w$.\\
Lemma~\ref{winstratexclusivity} is needed for the case of negation: \I have a winning strategy for $\mathbf{G}(\M,\mathbf{P}, \w: \neg \psi)$ iff \I have a winning strategy for $\mathbf{G}(\M,\mathbf{O}, \w: \psi)$. By the lemma, this is the case iff \I do not have a winning strategy for $\mathbf{G}(\M,\mathbf{P}, \w: \psi)$. By induction hypothesis, this is the case iff $\M, \w \not \models \psi$ iff $\M, \w \models \neg \psi$.
\end{proof}

We conclude this section with an important observation about the game for named models. Remember that in a named model, every world $\w$ has a name $i$, i.e. $\g(i)=\w$. Therefore, it is unambiguous, if we write $\mathbf{Q}, i:\phi$ for the game state $\mathbf{Q}, \w : \phi$. This, together with the fact that $\M \models R(i,j)$ iff $\g(i)\R\g(j)$ gives us the following equivalent formulations of the rules $(R_\Box)$, $(R_\Diamond)$ and $(R_U)$:

\begin{description}
\item[$(R_\Box)$] If $g=\mathbf{P}, i: \Box \psi$, then $g$ is labelled ``Y" and its immediate subtrees are $\mathbf{G}(\mathbf{P}, j: \neg R(i,j) \vee \psi)$, where $j$ ranges over the nominals.

\item[$(R_\Diamond)$] If $g=\mathbf{P}, i: \Diamond \psi$, then it is labelled ``I" and its immediate subtrees are $\mathbf{G}(\mathbf{P}, j: R(i,j) \wedge \psi)$, where $j$ ranges over the nominals.

\item[$(R_U)$] If $g=\mathbf{P}: \psi$, then it is labelled ``Y" and its immediate subtrees are $\mathbf{G}(\mathbf{P}, i: \psi)$, where $i$ ranges over the nominals.
\end{description}

From now on, we will tacitly assume that all models are named. This is justified by Lemma~\ref{namedlemma}. Also, we will only use the reformulations from above. We will see shortly the advantages of these reformulations.

\section{A Game for Validity}\label{disjgame}
In this section we develop a so-called \emph{disjunctive game} \cite{FMGiles} to model validity of a formula $\phi$. Intuitively, the idea is to play all possible games rooted in a game state $g$ at once. This is only possible using the reformulations of the rules $(R_\Box)$, $(R_\Diamond)$ and $(R_U)$: if we stick to those variants, the structure of the model does not affect the shape of the game tree anymore. To be more precise, if $\M_1$ and $\M_2$ are two models, then the game trees for $\mathbf{G}(\M_1,g)$ and $\mathbf{G}(\M_2,g)$ are identical, except maybe for the labelling of leaf states.

We have thus a uniform game tree and truth values become important only at the leafs. Consequently, we can design the disjunctive game as a two-player game. We might think of the two players as being the same as in the semantic game and correspondingly call them \Me and \You.
In order to compensate \Me for having to play ``blindly" in a game state $g$, i.e. without being guided by truth within a particular model, we allow \Me to add backtracking points $g\bigvee g$. The idea is that this allows \Me to first play on the left copy of $g$ and, in case of failure, change to right backup copy. In general, game states of the disjunctive game are thus of the form $D = g_1 \bigvee g_2 \bigvee .... \bigvee g_n$, where the $g_i$ are game states of the semantic game. According to our interpretation, such a disjunctive game state stands for the fact that for every model $\M$, \I have a winning strategy for $\mathbf{G}(\M,g_1)$ or for $\mathbf{G}(\M,g_2)$ ... or for $\mathbf{G}(\M,g_n)$\footnote{Unless noted otherwise, we identify a disjunctive state with the multiset of its game states. This implies that we consider two disjunctive states to be equal if they contain the same game states in the same numbers.}.

Another ingredient is necessary: to determine who is to move at a disjunctive game state $D$, we introduce \emph{regulation functions}. A regulation function $\rho$ maps non-elementary\footnote{A disjunctive state is called \emph{elementary} if all its game states are elementary, i.e. they involve only elementary formulas.} disjunctive states to one of its non-elementary game states $g_i$.

\begin{comment}
Informally, the disjunctive game proceeds as follows: Say, the current game state is $D\bigvee g$\footnote{If $D = g_1 \bigvee ... \bigvee g_n$, we conveniently write $D\bigvee g$ for the disjunctive state $g_1 \bigvee ... \bigvee g_n \bigvee g$} and the regulation function $\rho$ picks $g$. The player that is to move in $\mathbf{G}(g)$ makes the choice $g'$ and the game continues with $D\bigvee g'$. Alternatively, we allow \Me to add a backtracking point and \emph{duplicate} $g$: The game then continues with $D\bigvee g \bigvee g$. When the game reaches an elementary game state $g_1 \bigvee ... \bigvee g_n$, then \I win and \You lose if for every model $\M$, there is a $k$ such that \I win at $\mathbf{G}(\M,g_k)$. Otherwise, \I lose and \You win. 
\end{comment}

Formally, we identify the disjunctive game specified by the regulation $\rho$ and starting in the state $D$ with its game tree $\mathbf{DG}(D,\rho)$\footnote{If $\rho$ is clear from context (or not important), we conveniently write $\mathbf{DG}(D)$.}. It is defined recursively by specifying its immediate subtrees:
\begin{itemize}
\item If $\rho(D\bigvee g)=g$ and $g$ is labelled ``Y", then so is $D\bigvee g$. Its immediate subtrees are $\mathbf{DG}(D\bigvee g',\rho)$, where $g'$ ranges over all children of $g$ in $\mathbf{G}(g)$.
\item If $\rho(D\bigvee g) = g$ and $g$ is labelled ``I", then so is $D\bigvee g$. Its immediate subtrees are the following:
\begin{itemize}
    \item $\mathbf{DG}(D\bigvee g', \rho)$, where $g'$ ranges over all children of $g$ in $\mathbf{G}(g)$.
    \item The subtree $\mathbf{DG}(D\bigvee g \bigvee g)$\footnote{Due to this rule, the game may continue forever.}. We say that $D\bigvee g \bigvee g$ is obtained by \emph{duplicating} $g$.
\end{itemize}
    \item If $D=g_1 \bigvee ... \bigvee g_n$ is elementary, then it is labelled ``I", if for every model $\M$, there is some $i$ such that $g_i$ is labelled ``I" in $\mathbf{G}(\M,g_i)$. Otherwise, it is labelled ``Y".
\end{itemize}

\begin{example}
The simplest example to illustrate and explain the idea of disjunctive states is the game $\mathbf{DG}(\mathbf{P}:p\vee \neg p)$. Obviously, \I have a winning strategy for $\mathbf{G}(\M, \mathbf{P}:p \vee \neg p)$ for every model $\M$.

In the first round of $\mathbf{DG}(\mathbf{P}:p\vee \neg p)$, \You choose a nominal $i$ and the game proceeds with the state $\mathbf{P}, i: p \vee \neg p$  (\You made \Your choice according to the rule $(R_U)$). Now \I have two options: The first is to continue the game with either $\mathbf{P}, i: p$ or with $\mathbf{P}, i: \neg p$ (according to $(R_\vee)$). But both choices are bad: the first one, because there are models $\M$ with $\M, \g(i) \not \models p$ and the second, because there are models $\mathcal{N}$ with $\mathcal{N}, \g(i) \not \models \neg p$. This is where the second option comes in: \I duplicate the state $\mathbf{P}, i: p \vee \neg p$ and the game continues with $(\mathbf{P}, i: p \vee \neg p)\bigvee (\mathbf{P}, i: p \vee \neg p)$.

A regulation function will choose one of the two (say the left one). Now \I can choose to continue the game with $(\mathbf{P}, i : p) \bigvee (\mathbf{P}, i: p \vee \neg p)$. In the next round, the game continues with the left disjunct, and \I play $(\mathbf{P}, i : p) \bigvee (\mathbf{P}, i:\neg p)$. Finally, the game arrives at $(\mathbf{P}, i : p) \bigvee (\mathbf{O}, i: p)$ and \I win.
\end{example}

The rest of this section is dedicated to proving that this game indeed adequately models validity. Let us call a disjunctive state $D=g_1 \bigvee ... \bigvee g_n$ \emph{game valid}, if for every model $\M$, there is some $i$ such that \I have a winning strategy in $\mathbf{G}(\M,g_i)$. $D$ is called \emph{winning} if there is a regulation $\rho$ such that \I have a winning strategy for $\mathbf{DG}(D,\rho)$.

\begin{theorem}\label{disjgametheorem}
Every disjunctive state is game valid iff it is winning.
\end{theorem}

In light of Theorem~\ref{winningstrategy}, we immediately get the following proposition:

\begin{proposition}\label{validitycor}
The formula $\phi$ is valid iff $\mathbf{P}:\phi$ is winning.
\end{proposition}

Theorem~\ref{disjgametheorem} suffers from what Johan van Benthem calls ``$\exists$-sickness" \cite{vbenthbook}. However, our proof will offer a more constructive formulation: (1) There is a direct construction of winning strategies witnessing the game validity of $D$ from a winning strategy for \Me in $\mathbf{DG}(D)$. (2) There is a construction of a particular strategy $\sigma$ of \Me and a regulation $\rho$ with the following property: If $\sigma$ is not winning in $\mathbf{DG}(D,\rho)$, then one can extract a model $\M$ and winning strategies for \You for all $\mathbf{G}(\M,g)$, where $g \in D$.\footnote{If one changes the disjunctive game such that infinite runs are considered winning for \You, then one can extract the model and the strategies from \Your winning strategy in $\mathbf{DG}(D,\rho)$.}. We will now prove (1) and leave (2) for the next subsection.

\begin{proof}[of Theorem~\ref{disjgametheorem}, right-to-left]
Let $\sigma$ be a winning strategy for $\mathbf{DG}(D_0,\rho)$. We show by upwards-tree-induction on $\sigma$ that every $D$ appearing in $\sigma$ is game valid.

If $D$ is elementary, then it is valid, since $\sigma$ is winning. Let us deal with some exemplary cases of the inductive step. Let us start with two cases, where it is \My choice: If $D \bigvee g \bigvee g$ is valid and is obtained from $D\bigvee g$ by duplication of $g$, then clearly $D\bigvee g$ is valid too.

Let $D\bigvee (\mathbf{P}, i :\Diamond \psi)$ appear\footnote{In this proof, we conveniently write $D\bigvee g$ to indicate that $\rho(D\bigvee g)=g$.} in $\sigma$ and its child be $D\bigvee (\mathbf{P}, j:R(i,j) \wedge \psi)$. Let $\M$ be a model. If there is a state $g'$ of $D$ such that \I have a winning strategy in $\mathbf{G}(\M,g')$, then this state appears in $D\bigvee g$ and there is nothing to show. We will omit this trivial subcase in the other cases. If \I have a winning strategy $\mu$ for $\mathbf{G}(\M, \mathbf{P}, j:R(i,j) \wedge \psi)$, then \I have the following winning strategy for $\mathbf{G}(\M, \mathbf{P}, i: \Diamond \phi)$: choose the (world corresponding to the) nominal $j$ in the first round. The game then continues with $\mathbf{G}(\M, \mathbf{P}, j:R(i,j) \wedge \psi)$, where \I can use $\mu$ to win.

We check three cases where it is \Your move: if $D \bigvee (\mathbf{P}:\psi)$ appears in $\sigma$ then its children are of the form $D \bigvee (\mathbf{P}, i : \psi)$, where $i$ ranges over the nominals. If \I have a winning strategy $\mu_i$ for every $\mathbf{G}(\mathbf{P}, i:\psi)$, then, since $\M$ is named, \I have a winning strategy for $\mathbf{G}(\mathbf{P}:\psi)$: By the rule $(R_U)$, \You must choose a nominal $i$ in the first move. The game then proceeds with $\mathbf{G}(\mathbf{P}, i:\psi)$ and \I can use $\mu_i$ to win the game.

If $D\bigvee (\mathbf{P}, i: \psi_1 \wedge \psi_2)$ appears in $\sigma$, then its children are $D\bigvee (\mathbf{P}, i: \psi_1)$ and $D\bigvee (\mathbf{P}, i: \psi_2$). If $\M$ is a model such that \I have winning strategies $\mu_1$ for $\mathbf{G}(\M, \mathbf{P}, i:\psi_1)$ and  $\mu_2$ for $\mathbf{G}(\M, \mathbf{P}, i:\psi_2)$, then \I can use them to win  $\mathbf{G}(\M, \mathbf{P}, i:\psi_1\wedge \psi_2)$.

If $D\bigvee \mathbf{P}, i: \Box \psi$ appears in $\sigma$, then its children are $D\bigvee (\mathbf{P}, j: \neg R(i,j) \vee \psi)$, where $j$ ranges over the nominals. If \I have a winning strategy $\mu_i$ for every $\mathbf{G}(\M,\mathbf{P}, j: \neg R(i,j) \vee \psi)$, then \I can use them similarly to the above to win $\mathbf{G}(\M,\mathbf{P}, i: \Box \psi)$.
\end{proof}

\subsection*{The best way to play}
In this subsection we will construct a regulation $\rho_0$ and a strategy $\sigma$ for \Me for the game $\mathbf{DG}(D_0,\rho_0)$ such that $\sigma$ is guaranteed to be winning, if $D_0$ is game valid. Intuitively, the idea is as follows: In order to prevent bad choices, every time \I must move \I will use the duplication rule first. This way \I can always come back to have another shot. The trick is now to systematically exploit all possible choices. Playing this way ensures \Me to always take a good choice, if there is one. Here is an exact formulation of this construction:

\subsubsection*{Construction of the strategy $\sigma$}
Let $D^{el}$ be the disjunctive state obtained from $D$ by removing all non-elementary game states. Let $\#$ be an enumeration of triples $\langle D, g, i \rangle$, where $D$ is a non-elementary disjunctive state, $g$ is a non-elementary game state and $i$ is a nominal, and such that every triple appears in that enumeration infinitely often. We build a tree rooted in $D_0$ according to the following procedure:

\begin{enumerate}
\item Add $D_0$ as a root of the tree.
\item For every $n = \#(D,g,i)$, if $D\bigvee g$ is a leaf:
    \begin{enumerate}
    \item If \You move in $\mathbf{G}(g)$, then add as successors to $D\bigvee g$ all of \Your possible moves $D\bigvee g'$. For example, the successors of $D \bigvee (\mathbf{P}, i: \phi \wedge \psi)$ are $D\bigvee \mathbf{P}, i : \phi$ and $\mathbf{P}, i : \psi$. The successors of $D\bigvee (\mathbf{P}: \phi)$ are $D\bigvee  (\mathbf{P}, i: \phi)$, for every nominal $i$.
    \item If \I move in $\mathbf{G}(g)$ and $D^{el}$ is game valid, then add to $D \bigvee g$ an arbitrary move $D \bigvee g'$ of \Me.
    \item If \I move in $\mathbf{G}(g)$ and $D^{el}$ is not valid, then add to $D\bigvee g$ the child $D \bigvee g \bigvee g$. Afterwards, if there are only two options $g_1$ and $g_2$ for \Me in $\mathbf{G}(g)$, then add as a child $D\bigvee g_1 \bigvee g$ and as its child $D \bigvee g_1 \bigvee g_2$. For example, for $D\bigvee (\mathbf{P}, i: \phi \vee \psi)$, the tree looks as follows:
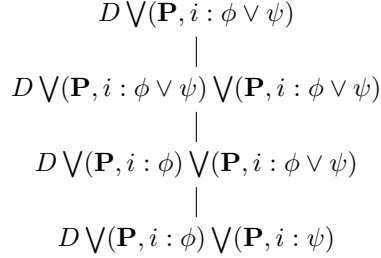
\begin{center} 
\begin{tikzpicture}[level distance = 1cm]
  			\node {\(D\bigvee (\mathbf{P}, i: \phi \vee \psi)\)}
  		  	child {node {\(D\bigvee (\mathbf{P}, i: \phi \vee \psi)\bigvee (\mathbf{P}, i: \phi \vee 					\psi)\)}
			child {node {\(D\bigvee (\mathbf{P}, i: \phi) \bigvee (\mathbf{P}, i: \phi \vee 									\psi)\)}
			child {node {\(D\bigvee (\mathbf{P}, i: \phi) \bigvee (\mathbf{P}, i: 	\psi)\)}}}};
		\end{tikzpicture}
\end{center}
    If there are infinitely many options in $\mathbf{G}(g)$, then they are parametrized by the nominals. For example, in $\mathbf{P}, j: \Diamond \phi$, \I must choose between $\mathbf{P}, k: R(i,k)\wedge \phi$, where $k$ ranges over the nominals. We first add $D\bigvee g \bigvee g$ as a child to $D\bigvee g$. Afterwards we add as a child $D\bigvee g \bigvee g(i)$, where $g(i)$ is the game state given by nominal $i$. In our example, for $D\bigvee (\mathbf{P},j:\Diamond \phi)$, we add $D\bigvee (\mathbf{P},i:\Diamond \phi) \bigvee (\mathbf{P},j:\Diamond \phi)$ and $D\bigvee (\mathbf{P},j:\Diamond \phi) \bigvee (\mathbf{P},i:R(i,j)\wedge \phi)$.
    \end{enumerate}
\end{enumerate}

The outline of the rest of the proof is as follows: If $\sigma$, as constructed above, is not a winning strategy, then there is at least one path $\pi$ through $\sigma$ rooted at $D_0$, such that $\pi$ does not end in a game valid elementary leaf. This means that $\pi$ either ends in a non-valid leaf or is infinite. We will now define a model $\M_\pi$ with the property that \I do not have a winning strategy for every $\mathbf{G}(\M_\pi, g)$, for any game state $g$ appearing along $\pi$\footnote{We say that a game state $g$ \emph{appears along} $\pi$, if it occurs as part of a disjunctive state in $\pi$.}. We call a model with this property a $\pi$-countermodel.

\begin{definition}
Let $D_\pi^n$ be the $n$-th node in the path $\pi$ and let us define the relation $i\sim_\pi^n j$ between two nominals $i$ and $j$ iff $\mathbf{O}, i: j$ or $\mathbf{O}, j: i$ appear in $D_\pi^n$. Let $\approx_\pi^n$ be the symmetric, reflexive and transitive closure of $\sim_\pi^n$. Let $i \approx_\pi j$, iff for some $n$, $i \approx_\pi^n j$. We write $[i]$ for the equivalence class of $i$. Since the elementary parts of the nodes accumulate in the course of $\pi$, we have for each $n$, $\approx_\pi ^n \subseteq \approx_\pi^{n+1}\subseteq \approx_\pi$ (if the length of $\pi$ is at least $n+1$).
We define the following named model $\M_\pi$:
\begin{itemize}
\item Worlds: Equivalence classes of nominals, 
\item Accessibility relation $\R_\pi$: We have $[i]\R_\pi[j]$ iff for some $i'\in [i]$ and $j'\in [j]$,  $k\in N$, $\mathbf{O}, k: R(i',j')$ appears along $\pi$.
\item Valuation function $\V_\pi$: $[i] \in \V_\pi(p)$ iff for some $i'\in [i]$, $\mathbf{O}, i': p$ appears in $\pi$.
\item Assignment $\g_\pi$: $\g_\pi(i)=[i]$.
\end{itemize}
\end{definition}

The following lemma shows that the definition of the equivalence relation is not arbitrary: In every $\pi$-countermodel of $D_\pi ^i$ equivalent nominals must name the same worlds.

\begin{lemma}\label{equcl}
Let $\M$ be a $\pi$-countermodel of $D_\pi^n$. Then $\M$ respects the equivalence $\approx_\pi^n$: If $i \approx^n_\pi j$, then $\g(i)=\g(j)$.
\end{lemma}

\begin{proof}
Let $i \sim_\pi^n j$. Then either $\mathbf{O}, i: j$ or $\mathbf{O}, j: i$ appear in $D_\pi^n$. Either way, to be a $\pi$-countermodel, $\M$ must satisfy $\g(i) = \g(j)$. The result for the symmetric, reflexive and transitive closure follows from the corresponding properties of $=$.
\end{proof}

\begin{lemma}\label{lemmamodel}
If $g$ appears along $\pi$, then \You have a winning strategy for $\mathbf{G}(\M_\pi, g)$.
\end{lemma}

\begin{proof}
We show the lemma by induction on $\mathbf{G}(g)$. The elementary cases where $g$ is of the form $\mathbf{O}, i: \phi$ are clear. Assume $g= \mathbf{P}, i: p$ appears along $\pi$, but $\M_\pi,[i]\models p$. The latter implies that for some $j \in [i]$, $\mathbf{O}, j: p$ appears along $\pi$. Since the elementary states of $\pi$ are accumulative, there is a disjunctive state $D$ in $\pi$ containing the three states $\mathbf{P}, i: p$, $\mathbf{O}, j: p$ and one of $\mathbf{O}, i: j$ or $\mathbf{O}, j: i$. Clearly, there is no model $\M$ satisfying $\M, \g(i) \not \models p$, $\M, \g(j) \models p$ and $\g(i)=\g(j)$ at the same time. Thus, \I could have easily won the game starting at $D$, a contradiction to the fact that $\sigma$ is not a winning strategy for \Me in $\mathbf{DG}(D,\rho)$. The cases for $\mathbf{P}, k: R(i,j)$ and $\mathbf{P}, i:j$ are similar.

Let us show some of the inductive steps: For the inductive step, suppose $g = \mathbf{P}, i: \psi_1 \vee \psi_2$ appears along $\pi$. By construction, both $\mathbf{O}, i:\psi_1$ and $\mathbf{O}, i:\psi_2$ appear along $\pi$. By inductive hypothesis, \You have winning strategies $\mu_1$ and $\mu_2$ for $\mathbf{G}(\M_\pi,\mathbf{P}, i: \psi_1)$ and $\mathbf{G}(\M_\pi,\mathbf{P}, i: \psi_2)$ respectively. \You can easily combine them to obtain a winning strategy for $\mathbf{G}(\M_\pi,\mathbf{P}, i: \psi_1 \vee \psi_2)$: in the first round \I choose to continue the game with $\mathbf{G}(\M_\pi, \mathbf{P}, i: \psi_k)$ and \You can use $\mu_k$ to win.

If $g=\mathbf{P}, i: \psi_1 \wedge \psi_2$ appears along $\pi$, then by construction, at least one of $\mathbf{P}, i: \psi_1$ and $\mathbf{P}, i: \psi_2$ appears along $\pi$. Without loss of generality, let $\mathbf{P}, i: \psi_1$ appear along $\pi$. By inductive hypothesis, \You have a winning strategy $\mu$ for $\mathbf{G}(\M_\pi, \mathbf{P}, i: \psi_1)$. Hence, \You can win $\mathbf{G}(\M_\pi,\mathbf{P},i:\psi_1 \wedge \psi_2)$ by choosing $\mathbf{P}, i: \psi_1$ in the first round and continuing along $\mu_1$.

If $\mathbf{P}:\psi$ appears along $\pi$, then by construction, there is a nominal $i$ such that $\mathbf{P}, i: \psi$ appears along $\pi$. By inductive hypothesis, \You have a winning strategy $\mu$ for $\mathbf{G}(\M_\pi,\mathbf{P}, i:\psi)$. But then \You can also win $\mathbf{G}(\M_\pi, \mathbf{P}, \psi)$ by choosing $i$ in the first round and continuing according to $\mu$.

If $\mathbf{P}, i: \Diamond \psi$ appears along $\pi$, then for every $j$, also $\mathbf{P}, j: R(i,j) \wedge \psi$ appears along $\pi$. By inductive hypothesis,\You have a winning strategy $\mu_j$ for $\mathbf{G}(\M_\pi,\mathbf{P}, j: R(i,j) \wedge \psi)$ for every $j$. Clearly, \You can combine them to a winning strategy for $\mathbf{G}(\M_\pi,i:\Diamond \psi)$.
\end{proof}

\begin{proof}[of Theorem~\ref{disjgametheorem}, left-to-right]
By contraposition: If $D_0$ is not winning, then, in particular, $\sigma$ from above is not a winning strategy for $\mathbf{DG}(D_0,\rho_0)$.  Let $\pi$ be a path through $\sigma$ rooted at $D_0$, such that $\pi$ does not end in a valid elementary leaf. By Lemma~\ref{lemmamodel}, \You have a winning strategy for  $\mathbf{G}(\M_\pi,g)$ for every $g$ appearing along $\pi$. In particular, \You have a winning strategy for all $\mathbf{G}(\M_\pi,g)$, where $g$ is in $D_0$. Hence, $D_0$ is not game valid.
\end{proof}

\section{From Strategies to Proof Systems}\label{proofsys}
Proposition~\ref{validitycor} shows that a regulation $\rho$ and a winning strategy for \Me in the game $\mathbf{DG}(\mathbf{P}:\phi)$ can be interpreted as a proof of $\phi$. However, due to the rules $(R_U)$, $(R_\Box)$ and $(R_\Diamond)$, this proof may be infinitely branching. In the following we will remedy this and give a finitized version of the disjunctive game.

\subsection{\Your optimal choices or how to achieve finite branching}
The idea is to use eigenvariables, similar to sequent calculi: for example, \I should be able to win $\mathbf{DG}(D \bigvee (\mathbf{P}, i: \Box \phi),\rho)$ iff \I can win $\mathbf{DG}(D \bigvee (\mathbf{P}, j: \neg R(i,j) \vee \phi),\rho)$, if $j$ stands for ``an arbitrary nominal". We may interpret this $j$ as \Your optimal choice. In technical terms, this translates into the condition that $j$ does not appear in the state $D\bigvee (\mathbf{P},i: \Box \phi)$.

\begin{proposition}\label{finite branching}
Let $D$ be any disjunctive state. 
\begin{enumerate}
\item $\mathbf{P}:\phi$ is winning iff $\mathbf{P},i:\phi$ is winning, where $i$ is a nominal not occurring in $\phi$.
\item $D \bigvee (\mathbf{P}, i: \Box \phi)$ is winning iff $D \bigvee (\mathbf{P}, j: \neg R(i,j) \vee \phi)$ is winning, where $j$ is a nominal not occurring in $D \bigvee (\mathbf{P}, i: \Box \phi)$. 
\item  $D\bigvee (\mathbf{O}, i: \Diamond \phi)$ is winning iff $D \bigvee (\mathbf{O}, j:  R(i,j) \wedge \phi)$ is winning, where $j$ is a nominal not occurring in  $D\bigvee (\mathbf{O}, i: \Diamond \phi)$.
\end{enumerate}
\end{proposition}

\begin{proof}[sketch]
Let us show 2: Let $\sigma$ be a winning strategy for \Me for the game  $\mathbf{DG}(D \bigvee (\mathbf{P}, j: \neg R(i,j) \vee \phi), \rho)$. Then the tree rooted in $D \bigvee (\mathbf{P}, i: \Box \phi)$ with $\sigma$ as its immediate subtree defines a winning strategy for \Me for $\mathbf{DG}(D \bigvee (\mathbf{P}, i: \Box \phi), \rho')$, where $\rho'$ is the same as $\rho$ except that $\rho'(D \bigvee (\mathbf{P}, i: \Box \phi))=\mathbf{P}, i: \Box \phi$.

If $D\bigvee (\mathbf{P}, i: \Box\phi)$ is not winning, then, by Theorem~\ref{disjgametheorem}, there is a model $\M = (\W,\R,\V,\g)$ such that \You have winning strategies for all $\mathbf{G}(\M,g)$ for $g\in D$ and for $\mathbf{G}(\M,\mathbf{P}, i:\Box \phi)$. In particular, \You have a winning strategy for $\mathbf{G}(\M,\mathbf{P},k: \neg R(i,k) \vee \phi)$ for some nominal $k$. Since truth of a formula in a model does not depend on nominals not appearing in that formula, we may assume that \You have winning strategies for all $\mathbf{G}(\M',g)$ for $g\in D$ and for $\mathbf{G}(\M',\mathbf{P}, i:\Box \phi)$. Here $\M'=(\W,\R,\V,\g')$ is the same as $\M$, except for $\g'$: it agrees with $\g$ on all nominals appearing in $D$ and $\phi$, but $\g'(j)=\g(k)$. Now \You have a winning strategy for $\mathbf{G}(\M',\mathbf{P},j:\neg R(i,j)\vee \phi)$. 
\end{proof}

\subsection{The proof system $\mathbf{DS}$}
We are now ready to formulate the sequent calculus $\mathbf{DS}$ (Figure~\ref{calculusfig}). We say that a string $i: \phi$ consisting of a hybrid logic formula $\phi$ and a nominal $i$ is a \emph{labelled formula} and we call an object $\Gamma \vdash \Delta$, where $\Gamma$ and $\Delta$ are multisets of formulas and labelled formulas, a \emph{sequent}. A disjunctive state $D$ can be rewritten as a sequent $\Gamma \vdash \Delta$ in the following way: $\Gamma$ comprises all (labeled) formulas with the prefix $\mathbf{O}$ in $D$ and $\Delta$ comprises all (labeled) formulas with the prefix $\mathbf{P}$ in $D$. For example, the disjunctive game state $\mathbf{O},i : \Box p \bigvee \mathbf{P}: p$ becomes the sequent $i:\Box p \vdash p$. There is thus a 1-1 correspondence between sequents and disjunctive states. 

Apart from the encoding of disjunctive states as sequents and the traditional bottom-up notation of proof trees, proofs in $\mathbf{DS}$ exactly correspond to \My winning strategies in the disjunctive game: the order of rule application defines a regulation function and determines \My moves in the strategy, branching in the proof tree corresponds to branching in the winning strategy, i.e. \Your possible moves. Infinitary branching is modified according to the discussion in the previous subsection. 
Duplication in the game takes the form of left and right contraction rules. The base rules are exactly the (encoding\footnote{We say that $\Gamma \vdash \Delta$ is \emph{elementary} iff its associated disjunctive state is. Similarly, $\Gamma \vdash \Delta$ is called \emph{valid}, if its associated disjunctive state is game valid.} of) valid disjunctive elementary states and thus winning for \Me. Using this correspondence we immediately get the following results:

\begin{proposition}
$\vdash \phi$ is provable in \(\mathbf{DS}\) iff $\mathbf{P}: \phi$ is winning.
\end{proposition}

\begin{theorem}
$\vdash \phi$ is provable in \(\mathbf{DS}\) iff $\models \phi$.
\end{theorem}

\begin{figure}

\begin{longtable}{l l}
\multicolumn{2}{l}{
        \textbf{Base rules}
        \medskip
        }
        \\*

 \multicolumn{2}{l}{
        {\hspace{5mm}
        \begin{prooftree}
        \hypo {}
        \infer1 [$(V)$ where \(\Gamma \vdash \Delta\) is elementary and valid]{\Gamma \vdash \Delta}
        \end{prooftree}
        }
        \bigskip
        }
        \\
  
   \multicolumn{2}{l}{
        \textbf{Structural Rules}
        \medskip
        }
        \\*
  
  {\hspace{5mm}
        \begin{prooftree}
        \hypo {\Gamma, i: \phi, i:\phi \vdash  \Delta}
        \infer1 [$(CL)$]{\Gamma, i:\phi \vdash  \Delta}
        \end{prooftree}
        \medskip
        }
		&
         {\begin{prooftree}
        \hypo {\Gamma \vdash  i:\phi,i:\phi,\Delta}
        \infer1 [$(CR)$]{\Gamma \vdash  i:\phi, \Delta}
        \end{prooftree}
        }
        \\

 {\hspace{5mm}
        \begin{prooftree}
        \hypo {\vdash  i:\phi}
        \infer1 [$(U)$]{\vdash \phi}
        \end{prooftree}
         \bigskip
        }
        \\

\multicolumn{2}{l}{
        \textbf{Logical connectives}
        \medskip
        }
        \\*
        
        \hspace{5mm}
        {\begin{prooftree}
        \hypo {\Gamma, i:\phi \vdash \Delta}
        \hypo{\Gamma, i:\psi \vdash \Delta}
        \infer2 [\((L_\vee)\)]{\Gamma, i:\phi \vee \psi\vdash \Delta}
        \end{prooftree}
        }
        \medskip
        &
        
        {\begin{prooftree}
        \hypo {\Gamma \vdash i: \phi, \Delta}
        \infer1[\((R_\vee^1)\)]{\Gamma \vdash i: \phi \vee \psi, \Delta}
        \end{prooftree}
        \medskip
        }
        
        \\
        
         \hspace{5mm}
        {\begin{prooftree}
        \hypo {\Gamma, i:\phi \vdash \Delta}
        \infer1 [\((L_\wedge^1)\)]{\Gamma, i:\phi \wedge \psi\vdash \Delta}
        \end{prooftree}
        }
        \medskip
        &
        
          {\begin{prooftree}
        \hypo {\Gamma \vdash i: \psi, \Delta}
        \infer1[\((R_\vee^2)\)]{\Gamma \vdash i: \phi \vee \psi, \Delta}
        \end{prooftree}
        }
        \medskip
        \\
	
    	\hspace{5mm}
    	 {\begin{prooftree}
        \hypo {\Gamma, i:\psi \vdash \Delta}
        \infer1 [\((L_\wedge^2)\)]{\Gamma, i:\phi \wedge \psi\vdash \Delta}
        \end{prooftree}
        }
        \medskip
        &
        
        {\begin{prooftree}
        \hypo {\Gamma \vdash i: \phi, \Delta}
        \hypo {\Gamma \vdash i: \psi, \Delta}
        \infer2[\((R_\wedge)\)]{\Gamma \vdash i: \phi \wedge \psi, \Delta}
        \end{prooftree}
        }
        \medskip
        \\ 
        
        \hspace{5mm}
          {\begin{prooftree}
        \hypo {\Gamma, i:\phi \vdash \Delta}
        \infer1 [\((R_\rightarrow^1)\)]{\Gamma\vdash i:\phi \rightarrow \psi, \Delta}
        \end{prooftree}
        }
        \medskip
        &
        
        {\begin{prooftree}
        \hypo {\Gamma \vdash i:\phi, \Delta}
        \hypo{\Gamma, i:\psi \vdash \Delta}
        \infer2 [\((L_\rightarrow)\)]{\Gamma, i:\phi \rightarrow \psi\vdash \Delta}
        \end{prooftree}
        }
        \medskip
        \\
        
        \hspace{5mm}
        {\begin{prooftree}
        \hypo {\Gamma \vdash i: \psi, \Delta}
        \infer1[\((R_\rightarrow^2)\)]{\Gamma \vdash i: \phi \rightarrow \psi, \Delta}
        \end{prooftree}
        }
        \medskip
        &
        
         {\begin{prooftree}
        \hypo {\Gamma \vdash, i:\phi, \Delta}
        \infer1 [\((L_\neg)\)]{\Gamma, i: \neg\phi\vdash \Delta}
        \end{prooftree}
        }
        \medskip
        \\
        
        \hspace{5mm}
         {\begin{prooftree}
        \hypo {\Gamma, i: \phi \vdash \Delta}
        \infer1[\((R_\neg)\)]{\Gamma \vdash i: \neg \phi, \Delta}
        \end{prooftree}
        }
        \bigskip
        \\
        
        \multicolumn{2}{l}{
        \textbf{Modal operators}
        \medskip
        }
        \\*

        \hspace{5mm}
         {\begin{prooftree}
        \hypo {\Gamma, j: \neg R(i,j) \vee \phi \vdash \Delta}
        \infer1 [\((L_\Box)\)]{\Gamma, i: \Box\phi\vdash \Delta}
        \end{prooftree}
        }
        \medskip
        
        &

        {\begin{prooftree}
        \hypo {\Gamma \vdash j: \neg R(i,j) \vee \phi, \Delta}
        \infer1 [\((R_\Box)\)]{\Gamma\vdash i: \Box\phi,\Delta}
        \end{prooftree}
        }
        \medskip
        
        \\
        
        \hspace{5mm}
         {\begin{prooftree}
        \hypo {\Gamma, j: R(i,j) \wedge \phi \vdash \Delta}
        \infer1 [\((L_\Diamond)\)]{\Gamma, i: \Diamond\phi\vdash \Delta}
        \end{prooftree}
        }
        \medskip
        
        &

        {\begin{prooftree}
        \hypo {\Gamma \vdash j: R(i,j) \wedge \phi, \Delta}
        \infer1 [\((R_\Diamond)\)]{\Gamma\vdash i: \Diamond\phi,\Delta}
        \end{prooftree}
        }
        \medskip
        
        \\
        
        \hspace{5mm}
         {\begin{prooftree}
        \hypo {\Gamma, j: \phi \vdash \Delta}
        \infer1 [\((L_@)\)]{\Gamma, i: @_j\phi\vdash \Delta}
        \end{prooftree}
        }
        
        &

        {\begin{prooftree}
        \hypo {\Gamma \vdash j: \phi, \Delta}
        \infer1 [\((R_@)\)]{\Gamma\vdash i: @_j\phi,\Delta}
        \end{prooftree}
        }
  
\end{longtable}

In the rules $(R_\Box)$, $(L_\Diamond)$ and $(U)$, the nominal $j$ must not occur in the lower sequent. This condition corresponds to \Your optimal choices. 
\caption{\label{calculusfig} The proof system $\mathbf{DS}$}
\end{figure}
The calculus $\mathbf{DS}$ takes a familiar form: The rules for the logical connectives are the (labeled) versions of the usual  propositional rules of sequent calculus for classical logic. The modal operator rules come in the form of their first-order translations. Apart from the axioms, $\mathbf{DS}$ can be therefore seen as a fragment of the usual sequent system for first-order logic. In turn, $\mathbf{DS}$ is an extension of the sequent calculus $\mathbf{G3K}$ \cite{NegriKripke} to hybrid logic. Similarly to $\mathbf{G3K}$, a failed proof search in $\mathbf{DS}$ directly gives rise to a countermodel. This follows from our proof of the left-to-right direction of Theorem~\ref{disjgametheorem}, where the the explicit countermodel $\mathcal{M}_\pi$ was constructed. 

\begin{example}
The rule

\begin{center}
\begin{prooftree}
 		\hypo {\Gamma, i:\phi \vdash i: \psi, \Delta}
        \infer1 [$(R_\rightarrow)$]{\Gamma\vdash i: \phi \rightarrow \psi,\Delta}
\end{prooftree}
\end{center}

is derivable. This can be seen using the rules $(R_\rightarrow^1)$, $(R_\rightarrow^2)$ and $(CR)$. Let us show how to prove \(i: \Box (p \wedge  q) \vdash i: \Box p \)  in \(\mathbf{DS}\):

\begin{center}
\begin{prooftree}
		\hypo{j: R(i,j) \vdash  j:  R(i,j)}
        \infer1[$(R_\neg)$]{\vdash j: \neg R(i,j),  j:  R(i,j)}
		\infer1[\((L_\neg)\)]{j: \neg R(i,j) \vdash j: \neg R(i,j)}
		\infer1[\((R_\vee^2)\)]{j: \neg R(i,j) \vdash j: \neg R(i,j) \vee p}
        \hypo{j:p \vdash j: p}
        \infer1[\((R_\vee^2)\)]{j:p \vdash j: \neg R(i,j) \vee p}
        \infer1[\((L_\wedge^1)\)]{j: p \wedge  q \vdash j: \neg R(i,j) \vee p}
		\infer2[\((L_\vee)\)]{j: \neg R(i,j) \vee (p \wedge  q) \vdash j: \neg R(i,j) \vee p}
        \infer1[\((L_\Box)\)]{i: \Box (p \wedge  q) \vdash j: \neg R(i,j) \vee p}
        \infer1[\((R_\Box)\)]{i: \Box (p \wedge  q) \vdash i: \Box p}
\end{prooftree}
\end{center}

The proof of \(i: \Box (p \wedge  q) \vdash i: \Box q \) looks similar. We can use both to give a proof of $\Box (p\wedge q) \rightarrow (\Box p \wedge \Box q)$:

\begin{center}
\begin{prooftree}
        \hypo{i: \Box (p \wedge  q) \vdash i: \Box p}
        \hypo{i: \Box (p \wedge  q) \vdash i: \Box q}
		\infer2[\((R_\wedge)\)]{i: \Box (p \wedge  q) \vdash i: \Box p \wedge \Box q}
 		\infer1[\((R_\rightarrow)\)] {\vdash i: \Box(p \wedge q) \rightarrow (\Box p \wedge \Box q)}
        \infer1 [\((R_U)\)]{\vdash \Box(p \wedge q) \rightarrow (\Box p \wedge \Box q)}
\end{prooftree}
\end{center}
\end{example}

\begin{remark}
The above example demonstrates the importance of the regulation function $\rho$ for the disjunctive game: the proof of $i: \Box (p \wedge q) \vdash i: \Box p$ relies on the possibility to expand the left-hand side before the right one. In game-theoretic terms, \I have a winning strategy for $\mathbf{DG}(D,\rho)$, where $D =\mathbf{O}, i: \Box (p \wedge q) \bigvee \mathbf{P}, i: \Box p$ and $\rho(D)=\mathbf{O}, i: \Box (p \wedge q)$. If, on the other hand, $\rho$ does never pick $\mathbf{O}, i: \Box (p\wedge q)$ (unless it has to), then \I do not have a winning strategy in the corresponding game. This complication does not arise in disjunctive games where the set of actions is finite, for example \cite{FMGiles}. Intuitively the idea there is for \Me to duplicate and exhaustively make one choice after the other. After finitely many rounds, all the possible actions have been played and \I can be sure to have played a good one, if there is any. It is clear that in general, this strategy does not work in the infinite case.
\end{remark}

\section{Conclusion and Future Work}
In this paper, we developed a semantic game for hybrid logic and proved its adequacy: \I have a winning strategy in the game of $\phi$ at world $\w$ over the model $\M$ iff $\M,\w\models \phi$. We proved that a version of the disjunctive game \cite{FMGiles} models validity in hybrid logic adequately.

In this paper, we have only looked at game-theoretic modelling of validity in basic hybrid logic. However, hybrid language allows us to characterize many classes of frames that lie beyond the expressivity of orthodox modal logic \cite{blackburn2002modal}. In the future, we plan to make use of this feature and present an interesting extension of the disjunctive game for hybrid logic to model validity over classes of frames characterizable in hybrid language and investigate the connections to Blackburn's dialogue game \cite{blackburn2002modal}.

Our aim, however, was not only to give game-theoretic characterizations of truth in a model and validity, but to do so in a way that allows for a natural lifting of the first to the second: the disjunctive game corresponds to simultaneously playing the semantic game over all possible models. The search for a winning strategy, in turn, can be straightforwardly formulated as an analytic calculus.

The lifting of semantic games to analytic calculi via a disjunctive game has been demonstrated on the propositional level for classical logic, \L ukasiewicz logic and Gödel logic \cite{FMGiles,FerLanPav20}. Our approach is the first to deal with infinitely many possible actions and we are keen to extend the known results to the first-order case of the above logics. Ultimately, the liftings rely on similar game-theoretic properties of the semantic and the corresponding disjunctive game. We plan to develop a powerful general lifting algorithm by exploiting, on an abstract level, the relevant features of these similarities.

\bibliographystyle{splncs04}
\bibliography{bibliography.bib}
\end{document}